# Identification of the high-pressure phases of $\alpha$-SnWO$_4$ combining x-ray diffraction and crystal structure prediction

Daniel Diaz-Anichtchenko[1,2], Jordi Ibáñez[2], Pablo Botella[1], Robert Oliva[2], Alexei Kuzmin[3], Li Wang[4], Yuwei Li[4], Alfonso Muñoz[5], Frederico Alabarse[6], Daniel Errandonea[1,*]

[1]Departamento de Física Aplicada-ICMUV, MALTA Consolider Team, Universidad de Valencia, Dr. Moliner 50, Burjassot, 46100 Valencia, Spain

[2]Geosciences Barcelona (GEO3BCN-CSIC), MALTA Consolider Team, Barcelona 08028, Spain

[3]Institute of Solid State Physics, University of Latvia, Kengaraga street 8, Riga LV-1063, Latvia

[4]School of Liberal Arts and Sciences, North China Institute of Aerospace Engineering, Langfang 065000, China

[5]Departamento de Física, MALTA-Consolider Team, Universidad de La Laguna, San Cristóbal de La Laguna, E-38200 Tenerife, Spain

[6]Elettra Sincrotrone Trieste, Trieste 34149, Italy

*Corresponding author: daniel.errandonea@uv.es

## Abstract

We have characterized the high-pressure behavior of $\alpha$-SnWO$_4$. The compound has been studied up to 30 GPa using a diamond-anvil cell and synchrotron powder X-ray diffraction. We report evidence of two structural phase transitions in the pressure range covered in our study, and we propose a crystal structure for the two high-pressure phases. The first one, observed around 12.9 GPa, has been obtained combining indexation using DICVOL and density-functional theory calculations. The second high-pressure phase, observed around 17.5 GPa, has been determined by using the CALYPSO code, the prediction of which was supported by a Le Bail fit to the experimental X-ray diffraction patterns. The proposed structural sequence involves two successive collapses of the unit-cell volume and an increase in the coordination number of Sn and W atoms. The room-temperature equations of state, the principal axes of compression and their compressibility, the elastic constants, and the elastic moduli are reported for $\alpha$-SnWO$_4$ and for the two high-pressure phases.

# 1. Introduction

SnWO₄ belongs to the AWO₄ orthotungstate family. This group of compounds is known for its interesting physical properties, including magnetism [1, 2], as well as for its wide range of applications which include gas sensing [3] and catalysis [4]. In addition, SnWO₄ exhibits a great potential as an anode material for lithium-ion batteries [5] and as a photocatalytic material [6, 7], among other applications. SnWO₄ has some peculiarities when compared with other orthotungstates which, depending on the ionic radius of the divalent cation A, mostly crystallize either in the monoclinic wolframite-type structure (space group $P2/c$) [8] or in the tetragonal scheelite-type structure (space group $I4_1/a$) [9]. In particular, the lone pair stereochemical activity of the $Sn^{2+}$ $s^2$ pair of valence electrons, that are not shared with another atom, causes a crystallographic distortion, which gives unique characteristics to the crystal structure of SnWO₄ [10]. In the structure of SnWO₄, Sn atoms are four-coordinated to oxygen atoms in a trigonal bipyramidal configuration with Sn in the vertex of the pyramid (see Fig. 1). So far, two polymorphs have been reported for SnWO₄ in the literature. One is orthorhombic (space group $Pnna$) [11], which is the focus of this study and will be referred to as α-SnWO₄ throughout this manuscript. The other displays a cubic structure (space group $P2_13$) and is designated as β-SnWO₄ in the literature [12].

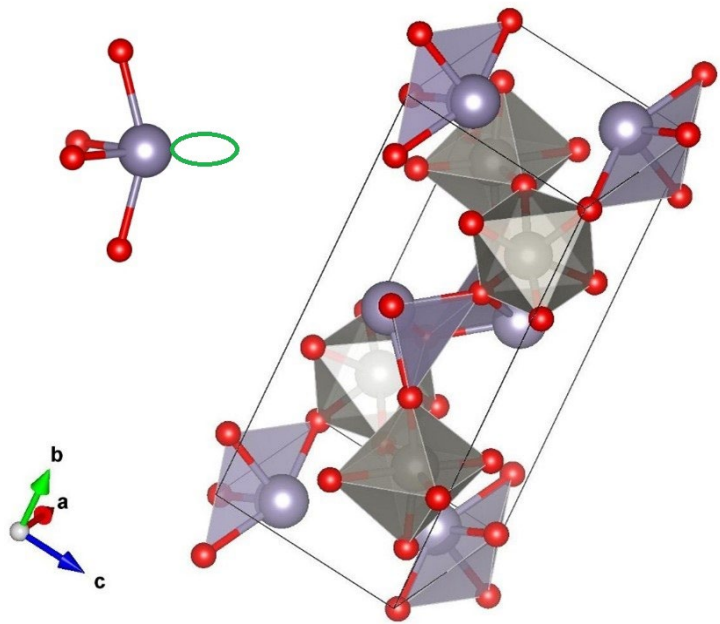

**Figure 1:** Perspective view of the crystal structure of α-SnWO₄. The SnO₄ coordination polyhedral units are shown in purple color and the WO₆ octahedral units are shown in grey color. The oxygen atoms are shown in red. The black solid lines represent the unit cell. The figure includes an isolated representation of the SnO₄ trigonal bipyramid with the lone electron pair represented as a green ellipse.

High pressure (HP) is an excellent tool to investigate materials. Pressure enables the significant alteration of atomic distances, thereby allowing for the adjustment of fundamental physical properties [13]. High pressure is especially valuable for studying compounds with atoms possessing a lone electron pair (LEP), resulting in structures with large areas of empty space and potential polyhedral tilting [14]. This is the case of $\alpha$-SnWO$_4$, which is formed by corner-linked distorted WO$_6$ octahedra and SnO$_4$ trigonal bipyramids with different degrees of distortion, with a zeolite-like channel running along the [001] direction and with the LEP of Sn pointing to the direction opposite of the base of the pyramid (see Fig. 1).

Over the past few decades, multiple research groups have investigated the family of orthotungstates under compression. [15, 16]. In wolframites, no evidence of phase transition has been found so far for any of these compounds up to pressures of approximately 20 GPa [17, 18, 19, 20]. In scheelites, phase transitions are found to occur approximately at 10 GPa [16]. Given the characteristics of the crystal structure of SnWO$_4$, and due to the LEP of Sn, a phase transition at relatively low pressure could be hypothesized for this compound. Indeed, theoretical studies proposed a phase transition from $\beta$-SnWO$_4$ to $\alpha$-SnWO$_4$ at around 2 GPa [20]. At the same time, the existence of an insulator-to-metal transition was theoretically predicted in $\alpha$-SnWO$_4$ at about 16 GPa [20] and observed experimentally above 5-7 GPa by mid-infrared spectroscopy [22]. Evidence of changes induced by pressure in the coordination polyhedra of $\alpha$-SnWO$_4$ has been determined from X-ray absorption studies (XAS) [20, 22]. Additional support for phase transitions in $\alpha$-SnWO$_4$ are reported in the literature [23]. Notably, the transition from $\beta$-SnWO$_4$ to $\alpha$-SnWO$_4$ has been observed at high temperatures [23]. This observation, considering the typical inverse relationship between temperature and pressure in crystal structures, supports the hypothesis that these phases may represent the ambient-pressure phase and its high-pressure equivalent. [24].

Up to date the crystal structure of the HP phase of $\alpha$-SnWO$_4$ remains unknown. In addition, a thorough investigation of the structural properties of this compound upon compression is still lacking. Hence, here we present a powder X-ray diffraction (XRD) study under HP conditions to investigate the behavior of $\alpha$-SnWO$_4$ under compression. We have found that this compound exhibits two first-order phase transitions, one around 12.9 GPa and a second one around 17.5 GPa. The resulting phases differ from the $\beta$ polymorph and imply important changes in the coordination polyhedra of Sn and W. To establish the crystal structure of the first HP phase, we have combined the analysis of XRD measurements with density-functional theory calculations, yielding a monoclinic structure which has been previously identified as a HP phase in other orthotungstates [25]. To determine the second HP phase, we utilized the Crystal structure AnaLYsis by Particle Swarm

Optimization (CALYPSO) code, which predicted a monoclinic structure that correctly matches the XRD patterns. This structure has been never reported before in any orthotungstate material.

## 2. Methodology

### 2.1 Experimental Details

Polycrystalline α-SnWO₄ was synthesized following the procedures outlined in References 12 and 26. Equimolar quantities of SnO (99.99%) and WO₃ (99.9%) powders were combined through mechanical mixing and subsequently sealed in a silica ampoule under vacuum conditions. The synthesis of SnWO₄ involved heating the ampoule to a temperature of 600 °C for a period of 8 hours, followed by a natural cooling process to room temperature (RT) along with the furnace. Laboratory powder XRD measurements performed with Cu K$_{\alpha 1}$ radiation showed that the process yielded the desired α-SnWO₄ phase. Weak peaks from other phases, which do not interfere with the present experiments, also show up in the scans. We could identify those peaks as one of the products of the preparation, WO₃.

A fine powder obtained from the synthesized material was employed for the high-pressure X-ray diffraction (HP-XRD) analysis. Angle-resolved high-pressure powder X-ray diffraction (HP-XRD) measurements were conducted at the Xpress beamline of the Elettra synchrotron, utilizing a monochromatic wavelength of 0.4956 Å and employing a PILATUS 3S 6M detector. The instrument was calibrated using LaB₆. The XRD patterns were obtained by integrating the two-dimensional diffraction rings from the detector using Dioptas [27]. The X-ray beam was focused down to a spot size of 50 μm × 50 μm. The experiments were conducted utilizing a membrane-driven diamond-anvil cell with diamond culets measuring 500 μm in diameter. We used in the process a stainless-steel gasket pre-indented to a thickness of 50 μm with a hole of 150 μm. The medium used for pressure transmission consisted of a mixture of ethanol, methanol, and water in a ratio of 16:3:1, which facilitates quasi-hydrostatic conditions at pressures reaching up to 10 GPa [28]. However, in oxides, it has been used to perform accurate studies up to 30 GPa [29]. Copper powder was included alongside the sample to serve as an internal standard for determining pressure [30]. The pressure was measured with an error smaller than 0.05 GPa for pressures smaller than 10 GPa and with an error smaller than 0.1 GPa for higher pressures.

### 2.2 Computational Details

First-principles calculations were conducted using density-functional theory (DFT) [31] and the projector-augmented wave (PAW) method [32, 33] within the Vienna ab-initio simulation package (VASP) [34]. A plane-wave energy threshold of 560 eV was implemented to guarantee precise outcomes. The exchange-correlation energy was described using the generalized-gradient approximation (GGA) with the Perdew-Burke-Ernzerhof for solids (PBEsol) functional [35]. The Monkhorst-Pack scheme [36] was employed to discretize the Brillouin zone (BZ) integrations with a 6 x 4 x 6 mesh in the irreducible BZ. In the relaxed equilibrium configuration, the forces on atoms were constrained to be below 1 meV/Å per atom along the Cartesian axes. For the calculations of the dynamical matrix using the direct force constant approach, precise force calculations are essential for obtaining accurate results [37]. We used a 2 x 2 x 2 supercell to obtain the phonon dispersion and check the dynamical stability of the different polymorphs. The mechanical characteristics of the different polymorphs of $SnWO_4$ were further analysed by determining the elastic constants through stress-strain calculations carried out in VASP with the Le Page [38] approach. The various elastic moduli were obtained from the elastic constants. The mechanical stability was assessed by applying the generalized Born stability criteria, taking into consideration the influence of hydrostatic pressure [38].

To predict the structure of the HP phases, we used first-principles energetic calculations and the CALYPSO methodology [39, 40]. CALYPSO is capable of rapidly determining the ground-state structures of materials exclusively from their chemical composition through the examination of the potential energy landscape. We searched for the HP structure of the post $\alpha$-$SnWO_4$ phase using simulation cell sizes up to 4 formula units at 0 K and 30 GPa. The initial step involved the generation of random structures that exhibit specific symmetries, employing crystallographic symmetry operations to derive atomic coordinates. The VASP code and the conjugate gradient method were utilized to carry out local optimizations until the relative energy changes drop below $10^{-5}$ eV per atom. The subsequent generation of structures was built by incorporating 60% of the structures initially generated, choosing those with the lower relative energies through Particle Swarm Optimization. An additional 40% of the structures in the new generation are randomly generated. The generation process employs a structural fingerprinting method known as bond characterization matrix, which explicitly prevents the occurrence of identical structures. These steps significantly enhance the diversity of structures, thereby improving the effectiveness of global structural exploration. After generating 1500 structures for each prediction, structural searching simulations stop. Then, the low-lying energy structures undergo further optimization using the procedure described in the previous paragraph. Our structure-simulation approach allowed us to explore numerous low-energy structural arrangements of $SnWO_4$ through the mimicking of structures of the $ABO_4$ system structures with A as a divalent metal element and B as either W or Mo, and the structures outlined in diagram proposed by Bastide [16]. They included a structure with eight atoms per formula unit usually known as BaWO₄-II (or PbWO₄-III).

## 3. Results and discussion

Figure 2 represents a selection of XRD scans measured for α-SnWO₄ at different pressures up to 12.7 GPa. The bottom part of the figure shows the pattern measured at 1.5 GPa together with the results from a Rietveld refinement. As can be seen in the figure, the observed reflections match well with the structure reported in the literature for the α phase [11]. The scans also exhibit a few weak features from impurity phases, mainly WO₃, which do not interfere with the signal from the orthorhombic α-SnWO₄ phase. Good agreement is also found between the cell parameters obtained from the Rietveld refinement with those in the bibliography and with our DFT calculations (see Table 1). The orthorhombic structure is found to be stable up to 12.7 GPa, where extra peaks start to show up in the scan. See for instance the peak identified with the symbol #.

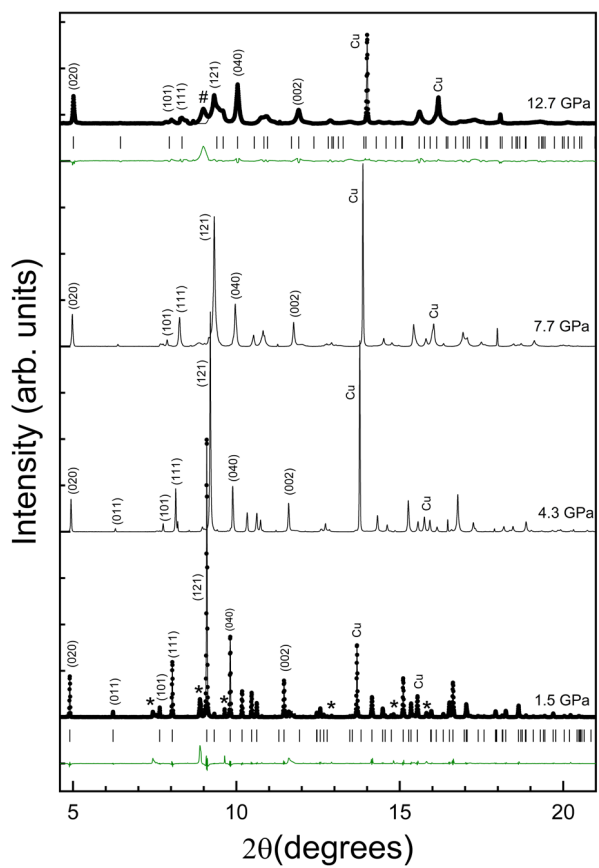

**Figure 2:** XRD patterns measured in α-SnWO₄ at selected pressures. The black symbols correspond to the experiments and the lowest and highest pressure. For these scans, the Rietveld refinement at 1.5 GPa (Le Bail fit at 12.7 GPa) is shown with a black line, and the corresponding residuals are plotted with a green line. The vertical ticks indicate the positions of calculated reflections. The peaks identified with asterisks are from impurities and can be assigned to WO₃ peaks. The Miller indexes for low-angle peaks are indicated. The Cu peaks used to determine pressure are identified.

**Table 1:** Unit-cell parameters and volume from the Rietveld refinements at 1.5 GPa pressure. They are compared with results from present calculations and previous experiments, which are reported at ambient pressure [11].

|  | Exp. | Ref. 11 | DFT |
|---|---|---|---|
| a (Å) | 5.5905(5) | 5.6270(3) | 5.5979 |
| b (Å) | 11.592(1) | 11.6486(7) | 11.7201 |
| c (Å) | 4.9653(4) | 4.9973(3) | 4.9926 |
| V (Å³) | 321.78(5) | 327.56(3) | 327.55 |

A gradual broadening of the peaks with increasing pressure is observed in the α-polymorph. At the same pressure, the XRD pattern becomes spotty due to recrystallization and the appearance of preferred crystallite orientations. It is mandatory to comment that the peak refinement cannot be completely successful in such cases due to the non-symmetry of the peak shape, which may be related to the quality of the powder used and, to the non-statistical goodness of the powder, which causes this asymmetry. The consequence is that the fit cannot be very accurate, even though the cell parameters are well adjusted. Another observation in the α-phase is the merge of the peaks, as can be seen in Fig. 2, which is a consequence of the anisotropic compressibility, which will be discussed in detail below.

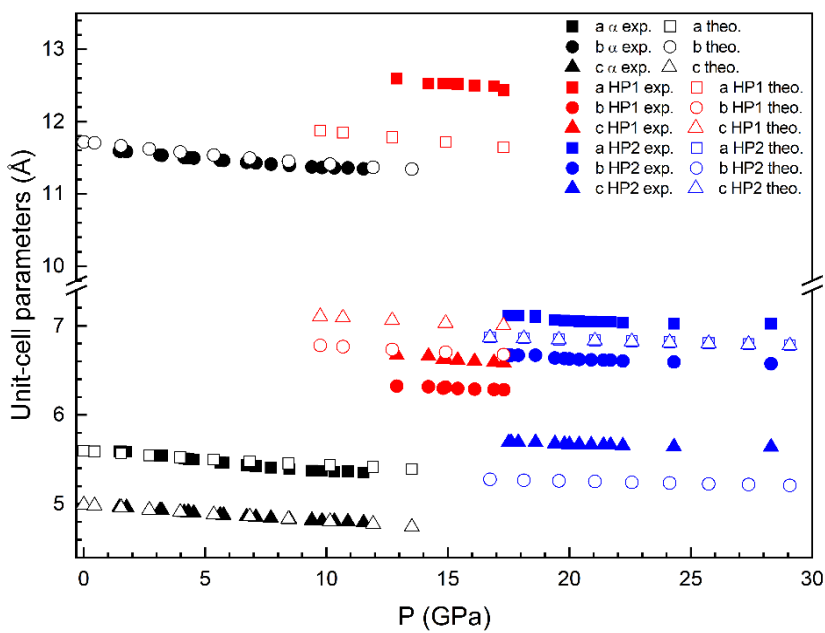

**Figure 3:** Pressure dependence of *a*, *b*, and *c* lattice parameters of α phase obtained from experiments and computer simulations, together with the unit-cell parameters of the two high-pressure phases, HP1 and HP2, reported in this work.

Figure 3 shows the pressure dependence of the unit-cell parameters for the α phase as obtained with Le Bail fits, together with the results of DFT calculations. As can be seen in the figure, a good agreement is found between experimental and theoretical results, with an overall difference as low as ~0.6 % and even lower (see Table 1). To further illustrate the good agreement between the XRD analysis for the α phase and the DFT calculations, we show in Table 2 the atomic coordinates obtained from experiments and calculations (at 0 GPa for the DFT calculations and at 1.5 GPa for the Rietveld refinement). There is excellent agreement between both collections of data, considering the different pressure values in both cases.

**Table 2:** Atomic positions of the α structure determined from experiments (Exp.) at 1.5 GPa and from DFT calculations (Theo.) at 0 GPa.

| Atom | Site | Exp. | Theo. |
|------|------|------|-------|
| Sn | 4c | (0.25, 0, 0.2196(4)) | (0.25, 0, 0.2202) |
| W | 4d | (0.6677, 0.25, 0.25) | (0.6515, 0.25, 0.25) |
| $O_1$ | 8e | (0.3765(9), 0.2013(9), 0.5012(9)) | (0.3652, 0.2021, 0.5017) |
| $O_2$ | 8e | (0.1019(9), 0.1039(9), 0.8963(9)) | (0.0931, 0.1039, 0.9012) |

The pressure-volume data was analyzed by fitting them with a third-order Birch–Murnaghan equation of state [41] using the EoSFit program [42]. The experimental data and the fitted EOS are represented in Fig. 4. The obtained zero-pressure bulk modulus ($B_0$), its pressure derivative ($B_0'$), and the volume ($V_0$) at zero pressure are given in Table 3. No experimental bulk modulus has been reported in the bibliography for α-SnWO$_4$ so far, but the bulk modulus and its pressure derivative determined from our experiments are slightly smaller than the theoretical values (see Table 3). Interestingly, the resulting bulk modulus obtained for α-SnWO$_4$ is significantly lower than that reported for wolframite-type compounds, which have $B_0$ values between 123 and 160 GPa [11,43]. In contrast, the bulk modulus of α-SnWO$_4$ is comparable to that of scheelite-type compounds, with $B_0$ values ranging from 60 and 75 GPa [16]. This could potentially explain why no phase transitions are observed in the less compressible wolframites, while they are found to occur in the more compressible tungstates such as the scheelites and, as will become evident next, in α-SnWO$_4$.

**Table 3:** The unit-cell volume ($V_0$), bulk modulus ($B_0$), and bulk modulus pressure derivative ($B_0'$) at zero pressure determined using a third-order Birch–Murnaghan EOS. We present results from experiments and calculations (our calculations and those of Ref. [21]).

| | $V_0$ (Å$^3$) | $B_0$ (GPa) | $B_0'$ |
|---|---|---|---|
| α-SnWO$_4$ (Our experiments) | 327.6 | 73.6 ± 5.6 | 3.4 ± 1.5 |
| α-SnWO$_4$ (Our calculations) | 327.6 | 87.9 | 4.0 |
| α-SnWO$_4$ Ref. [21] | 326.4 | 85.1 | 4.5 |

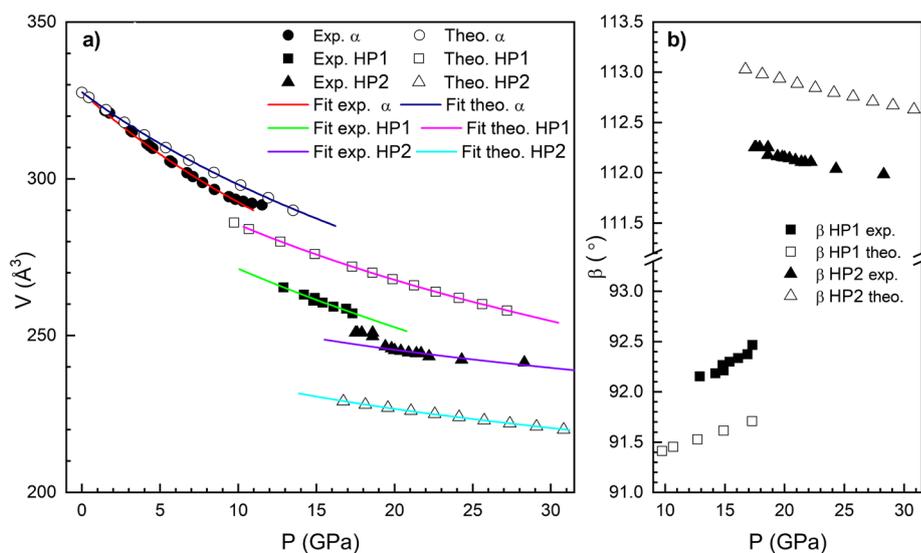

**Figure 4:** Pressure dependence of the volume $V$ of the phases α, HP1, and HP2. The lines are the least squares fits. The volume of HP1 is represented as $V/2$, because it contains twice the formula units as the other two phases.

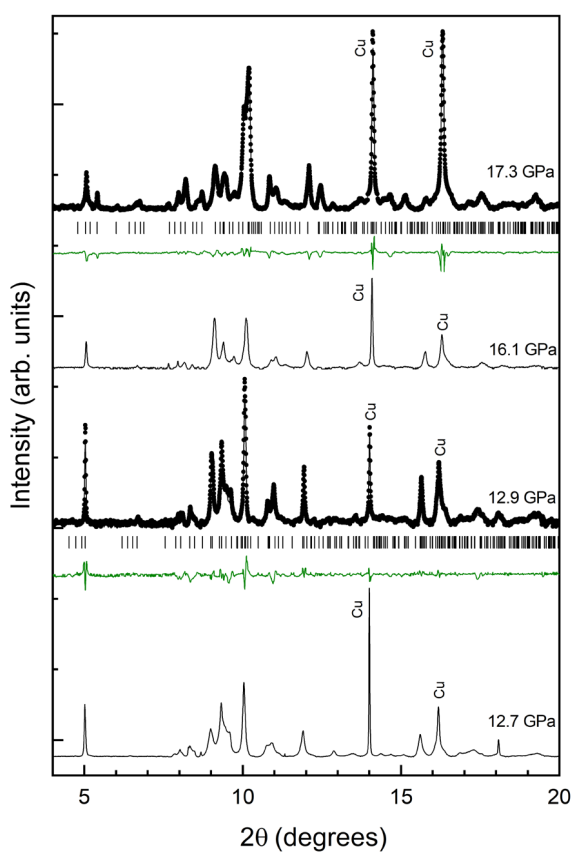

**Figure 5:** XRD patterns measured in α-SnWO4 at 12.7 GPa and the first HP phase of SnWO$_4$ (HP1 phase throughout the manuscript) at selected pressures from 12.9 to 17.3 GPa. At 12.9 and 17.3 GPa, the patterns are shown with black symbols, Le Bail fits with black lines, and residuals with green lines. The vertical ticks are the calculated positions of reflections. The Cu peaks used to determine pressure are identified in all the scans.

In Fig. 5 we show XRD patterns measured from 12.7 GPa to 17.3 GPa. The changes in the XRD pattern from 12.7 to 12.9 GPa are noticeable. The presence of a phase transition is signified by them. The resulting HP phase (denominated as HP1 from now on) remains stable up to 17.3 GPa. We found that XRD patterns of HP1 cannot be assigned to β-SnWO₄. Notice that the structure of β-SnWO₄ is cubic and consequently has fewer reflections than α-SnWO₄, which is orthorhombic. However, the number of peaks in our XRD patterns increases from 12.7 to 12.9 GPa, thus suggesting a decrease in the symmetry for phase HP1.

An indexation using DICVOL of the peaks observed below 12° at 12.9 GPa gives as the unit-cell with the best figure of merit a monoclinic one described by space group $P2_1/n$ with unit-cell parameters that resemble those of the BaWO₄-II type structure [44]. This unit cell contains eight formula units; therefore, it is not within the predictions of CALYPSO (see below). We have simulated this monoclinic phase using DFT, taking as the starting model the isomorphic structure reported for BaWO₄ by Kawada *et al.* [44] but substituting Ba with Sn. We found that the enthalpy of this structure becomes smaller than that of the $\alpha$ phase at 16 GPa, which supports that the BaWO₄-II-type structure is a good candidate for the HP1 phase. The results of enthalpy calculations are shown in Fig. 6. They show that from 0 to 16 GPa $\alpha$-SnWO₄ is the phase with the lowest enthalpy. i.e. the thermodynamically most stable phase. Our calculations of the phonon dispersion, as illustrated in Figure S1 in the Supplementary Information, further reinforce the stability of this phase by demonstrating that all phonon branches exhibit positive values. Phonon dispersion calculations for the HP1 also showed that this phase is dynamically stable at pressures where it was observed (See Fig. S2 in the Supplementary Information). Figure 6 includes, in addition to $\alpha$-SnWO₄ and HP1, structures proposed by CALYPSO with the lowest enthalpies, named HP2, HP3, HP4, HP5, and HP6, which will be discussed later.

Le Bail fits assuming the HP1 structure are in good agreement with the XRD patterns measured from 12.9 to 17.3 GPa, as shown in Fig. 5. Thus, both our XRD experiments and DFT calculations support that the BaWO₄-II type structure (HP1) is the first HP structure of α-SnWO₄. The unit-cell parameters we determined from experiments for HP1 at 12.9 GPa are $a$ = 12.593(6) Å, $b$ = 6.324(3) Å, $c$ = 6.670(3) Å, and $\beta$ = 92.17(9)°. The calculated atomic positions at the same pressure are reported in Table 4. It is not surprising to find HP1 as a HP structure of α-SnWO₄. This structure has been observed in other tungstates [45] and it is likely to occur according to the crystal chemistry arguments elaborated by Bastide and commonly used to predict HP structures of oxides [16]. The crystal structure of the monoclinic phase HP1 is shown in Fig. 7. The structure contains SnO₉ polyhedra interconnected by edge and corner with WO₆ octahedra. Notably, the phase transition involves an increase in the coordination number of Sn atoms, but not in the W atoms. The change in the coordination of Sn atoms is a consequence of the suppression of the LEP and the formation of extra bonds with second neighboring oxygen atoms, a common HP phenomenon in compounds with atoms with a LEP at ambient pressure [14]. In phase HP1, both polyhedral units display considerable distortion.

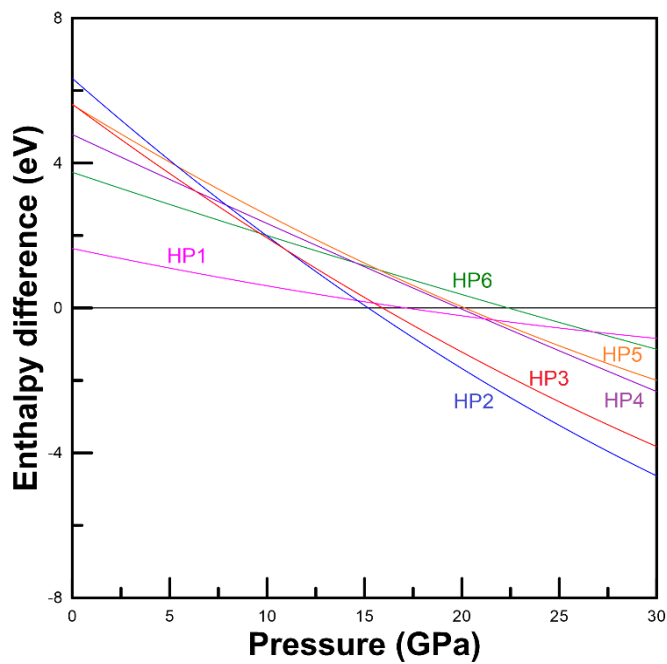

**Figure 6:** Pressure dependence of enthalpy difference with respect to α-SnWO₄ for all structures predicted by CALYPSO and the BaWO₄-II-type (HP1) structure.

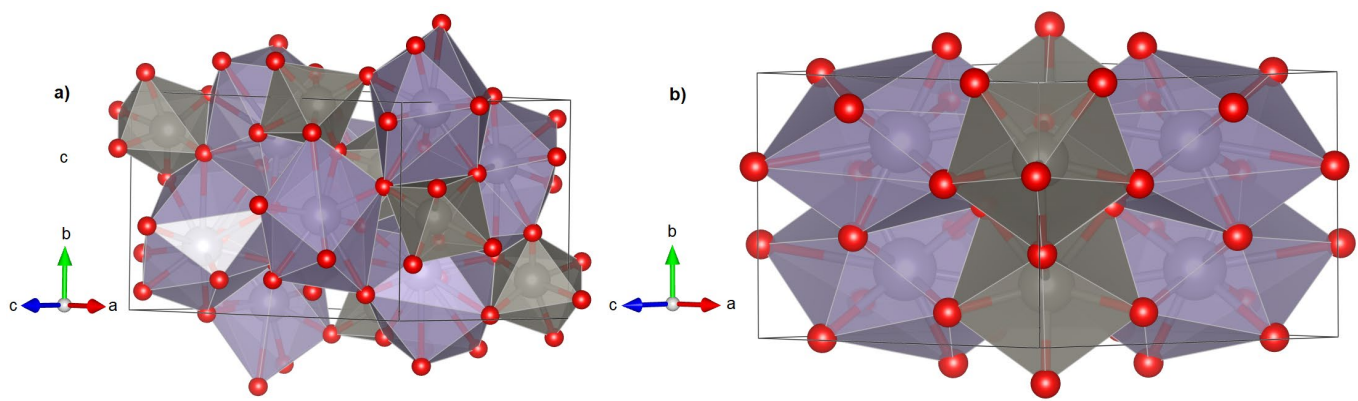

**Figure 7:** Schematic representation of phases (a) HP1 and (b) HP2 of SnWO₄. Sn coordination polyhedra are shown in purple and W coordination polyhedra in grey. The oxygen atoms are shown in red. The black solid lines represent the unit cell.

**Table 4:** DFT calculated atomic positions of the HP1 (BaWO$_4$-II-type) structure at 12.7 GPa.

| Atom | Site | Atomic coordinates |
|------|------|--------------------|
| Sn$_1$ | 4e | (0.33829, 0.79793, 0.66962) |
| Sn$_2$ | 4e | (0.36386, 0.53593, 0.11084) |
| W$_1$ | 4e | (0.41355, 0.3277, 0.57532) |
| W$_2$ | 4e | (0.40048, 0.02982, 0.1424) |
| O$_1$ | 4e | (0.38754, 0.48703, 0.78093) |
| O$_2$ | 4e | (0.29723, 0.89683, 0.27521) |
| O$_3$ | 4e | (0.45314, 0.83364, 0.97346) |
| O$_4$ | 4e | (0.26949, 0.22667, 0.56594) |
| O$_5$ | 4e | (0.42827, 0.2458, 0.30723) |
| O$_6$ | 4e | (0.31252, 0.15357, 0.96983) |
| O$_7$ | 4e | (0.47803, 0.10577, 0.68002) |
| O$_8$ | 4e | (0.41446, 0.59134, 0.44285) |

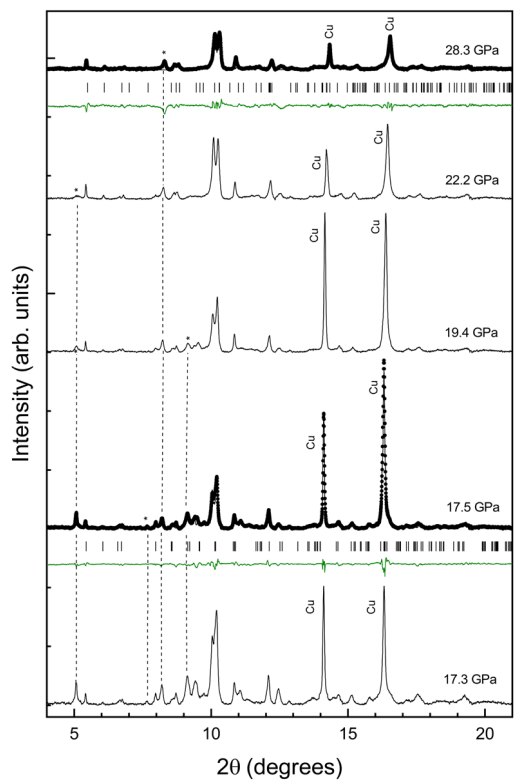

**Figure 8:** XRD patterns measured in SnWO$_4$ at selected pressures from 17.3 to 28.3 GPa. At 17.5 and 28.3 GPa the black symbols correspond to the experimental results. For these scans, Le Bail refinements assuming the HP2 phase are shown with black lines, and the corresponding residuals are plotted with green lines. The peaks with asterisks are from the coexisting HP1 phase peaks. We follow with dashed of peaks from HP1 when pressure increases.

In Fig. 8 we present XRD patterns measured from 17.3 to 28.3 GPa. From 17.3 to 17.5 GPa there are changes indicating the occurrence of a second phase transition. We have found that the XRD patterns measured from 17.5 to 28.3 GPa can be explained by one of the candidate structures predicted by CALYPSO, which we named HP2. CALYPSO also found other structures which were thermodynamically competitive with HP1 and HP2, but which are not compatible with the XRD patterns of the second HP phase. For the sake of completeness, the crystallographic information of these structures, named HP3, HP4, HP5, and HP6, can be found in the Supplementary Material. Fig. 8 shows Le Bail fits to the XRD patterns measured at 17.5 and 28.3 GPa assuming the HP2 structure. The quality of the fits supports the predictions of CALYPSO. HP1 and HP2 coexist up to 28.3 GPa but the peaks of HP1 gradually lose intensity as pressure increases. The residual peaks of phase HP1 are identified with asterisks and followed with dashed lines in Fig. 8.

HP2 is a monoclinic structure described by space group $P2_1/n$. It has four formula units per unit cell. The unit-cell parameters determined from the Le Bail fit at 17.5 GPa for phase HP2 are $a = 7.111(6)$ Å, $b = 6.670(6)$ Å, $c = 5.692(6)$ Å, and $\beta = 112.25(15)^o$. The calculated atomic positions of phase HP2 at 18.1 GPa are given in Table 5. The stability of phase HP2 is supported by our phonon dispersion calculations which show that all phonon branches of this structure are positive (See Fig. S3 in the Supplementary Information). The crystal structure of the monoclinic phase HP2 is shown in Fig. 7. The structure contains $SnO_{10}$ polyhedra interconnected by edge and corner with other $SnO_{10}$ polyhedra and $WO_7$ polyhedra. In this case, the phase transition involves an increase in the coordination number of both cations. On decompression from 28.3 GPa, we recovered a mixture of phases HP1 and HP2 showing that the bond formation induced by pressure is an irreversible process.

**Table 5:** DFT calculated atomic positions of the HP2 structure at 18.1 GPa.

| Atom | Site | Atomic positions |
|------|------|------------------|
| $Sn_1$ | 4e | (0.36598, 0.75, 0.86608) |
| $W_1$ | 4e | (0.91120, 0.70409, 0.91122) |
| $O_1$ | 4e | (0.32791, 0.15983, 0.32800) |
| $O_2$ | 4e | (0.31561, 0.39286, 0.97882) |
| $O_3$ | 4e | (0.89001, 0.99998, 0.11004) |
| $O_4$ | 4e | (0.02118, 0.60703, 0.68448) |

We display the pressure dependence of the unit-cell parameters and volume of phases HP1 and HP2 in Fig. 4. The volume of phase HP1 has been divided by 2 to allow a direct comparison with the other phases. Experiments and calculations show a qualitatively similar behavior. According to experiments the phase transitions involve two volume discontinuities of 7% and 3%, respectively. According to calculations, the discontinuities are 3% and 15%, respectively. The pressure-volume data were fitted with a third-order Birch–Murnaghan equation of state. For phase HP1 the following EOS parameters were extracted from experiments: $V_0$ = 592.3 Å$^3$, $B_0$ = 100.9 GPa and $B_0'$ = 2.9 and from calculations $V_0$ = 613.4 Å$^3$, $B_0$ = 92.1 GPa and $B_0'$ = 4.68. As expected, the bulk modulus of the high-pressure phase is higher than that of the ambient pressure phase. For HP2 we obtained from experiments $V_0$ = 266.7 Å$^3$, $B_0$ = 179.9 GPa and $B_0'$ = 4.9; and from calculations $V_0$ = 243.6 Å$^3$, $B_0$ = 183.6 GPa and $B_0'$ = 5.5. The second transition involves a large decrease of the compressibility.

Since HP1 and HP2 are monoclinic, their primary compressibility axes need to be identified by analyzing the eigenvalues and eigenvectors of the compressibility tensor, which were calculated through PASCal [46]. The results are shown in Table 6 and compared with the axial compressibilities of orthorhombic α-SnWO$_4$. The behavior of this phase is highly anisotropic, with the least compressible direction along the c-axis. In phase HP1 the compressibility is less anisotropic than in the α-phase, with the (4,0,9) direction being slightly more compressible than the other two directions. In phase HP2 the compressibility is highly anisotropic, with the (10,0,1) direction being the most compressible axis. For this phase, a remarkably low compressibility is obtained for the (1,0,3) direction, which explains the relatively large modulus obtained for HP2 ($B_0$ = 179.9 GPa).

**Table 6.** Experimental linear compressibilities α-SnWO$_4$ and linear compressibilities and main axis of compressibility for phases HP1 and HP2.

| α-SnWO$_4$ | HP1 | | HP2 | |
|---|---|---|---|---|
| $\kappa_a$ = 5.2(2) 10$^{-3}$ GPa$^{-1}$ | $\kappa_1$ = 3.3(2) 10$^{-3}$ GPa$^{-1}$ | $e_{v1}$ ≈ (4,0,9) | $\kappa_1$ = 3(3) 10$^{-3}$ GPa$^{-1}$ | $e_{v1}$ ≈ (10,0,1) |
| $\kappa_b$ = 4.00(4) 10$^{-3}$ GPa$^{-1}$ | $\kappa_2$ = 2.8(2) 10$^{-3}$ GPa$^{-1}$ | $e_{v2}$ ≈ (11,0,17) | $\kappa_2$ = 0.9(10) 10$^{-3}$ GPa$^{-1}$ | $e_{v2}$ ≈ (1,0,3) |
| $\kappa_c$ = 2.50(7) 10$^{-3}$ GPa$^{-1}$ | $\kappa_3$ = 2.7(1) 10$^{-3}$ GPa$^{-1}$ | $e_{v3}$ ≈ (0,1,0) | $\kappa_3$ = 1.7(1) 10$^{-3}$ GPa$^{-1}$ | $e_{v3}$ ≈ (0,1,0) |

Finally, we have also calculated the elastic constants for the three phases of SnWO$_4$ observed in this work. The results are summarized in Table 7. The derived elastic constants meet the stability criteria outlined in the generalized Born theory [47], indicating that the three structures are not only dynamically stable, but also mechanically stable. From the calculated elastic constants, we have obtained the bulk modulus (B), the shear modulus (G), the modulus of Young (E), and ratio of Poisson (ν) of the three phases. These elastic moduli have been obtained using the Hill approximation [49]. The calculated values are also summarized in Table 7. For α-SnWO$_4$, the obtained bulk modulus, 84.9

GPa, is in between the value obtained from experiments, 73.6 ± 5.6 GPa, and the value obtained with DFT calculations from the energy-volume results, 87.9 GPa. On the other hand, the value of the Young's modulus (104.6 GPa) is 23% smaller than the bulk modulus. This means that the resistance of α-SnWO₄ to tensile or compressive stress exceeds the resistance to volumetric compression. Conversely, the shear modulus of 40.4 GPa is notably lower than the bulk modulus, suggesting a preference for shear deformations rather than volume contraction in α-SnWO₄. For HP1, according to the present calculations, the bulk modulus at 16.7 GPa is 157.4 GPa. This value compares well with the value obtained from the third-order EOS reported above (170.2 GPa). In this case, the Young´s modulus is 17% smaller than the bulk modulus, which implies that volumetric compression requires larger stresses than lengthwise deformations. For HP2, from the calculated elastic constants, the bulk modulus at 22.6 GPa is 334.6 GPa, which agrees within 10% with the value estimated from the third-order EOS reported above (307.9 GPa). As in HP1, the Young´s modulus of HP2 is smaller than the bulk modulus (13%). Regarding the shear modulus of H1 and HP2, as occurs in α-SnWO₄, it is much smaller than the other moduli (48.4 and 109.2 GPa for HP1 and HP2, respectively). The three phases discussed exhibit a natural resistance to shear deformation that is considerably lower than their stiffness in response to uniaxial stress or hydrostatic pressure. According to the present results, the Poisson's ratios of the three phases are between 0.295 and 0.361, which are typical values for solids. It is evident that the interatomic bonding forces are mainly central in the three phases, with ionic bonding being more dominant than covalent bonding [49], which agrees with the conclusions from computer simulations reported by Stoltzfus *et al*. [50].

**Table 7.** Calculated elastic constants for α-SnWO₄ at 0 GPa, HP1 at 16.7 GPa, and HP2 at 22.6 GPa. The elastic moduli obtained from them are also included.

| α-SnWO₄ | | | | |
|---|---|---|---|---|
| $C_{11}$ = 157.5 GPa | $C_{22}$ = 171.8 GPa | $C_{33}$ = 126.5 GPa | $C_{44}$ = 59.9 GPa | $C_{55}$ = 27.1 GPa |
| $C_{66}$ = 27.6 GPa | $C_{12}$ = 42.6 GPa | $C_{13}$ = 53.7 GPa | $C_{23}$ = 59.4 GPa | |
| B = 84.9 GPa | G = 40.4 GPa | E = 104.6 GPa | $\nu$ = 0.295 GPa | |
| HP1 | | | | |
| $C_{11}$ = 194.9 GPa | $C_{22}$ = 231.8 GPa | $C_{33}$ = 229.6 GPa | $C_{44}$ = 54.2 GPa | $C_{55}$ = 50.5 GPa |
| $C_{66}$ = 50.2 GPa | $C_{12}$ = 126.2 GPa | $C_{13}$ = 131.2 GPa | $C_{23}$ = 59.4 GPa | $C_{15}$ = 9.1 GPa |
| $C_{25}$ = 3.0 GPa | $C_{35}$ = 7.9 GPa | $C_{46}$ = -8.7 GPa | | |
| B = 157.4 GPa | G = 48.4 GPa | E = 131.7 GPa | $\nu$ = 0.361 GPa | |
| HP2 | | | | |
| $C_{11}$ = 420.4 GPa | $C_{22}$ = 461.3 GPa | $C_{33}$ = 477.6 GPa | $C_{44}$ = 135.1 GPa | $C_{55}$ = 148.5 GPa |
| $C_{66}$ = 131.7 GPa | $C_{12}$ = 254.0 GPa | $C_{13}$ = 306.3 GPa | $C_{23}$ = 283.4 GPa | $C_{15}$ = -7.6 GPa |
| $C_{25}$ = 31.9 GPa | $C_{35}$ = 72.0 GPa | $C_{46}$ = 3.1 GPa | | |
| B = 334.6 GPa | G = 109.2 GPa | E = 295.4 GPa | $\nu$ = 0.353 GPa | |

## 4. Conclusions

High-pressure synchrotron powder X-ray diffraction measurements up to 30 GPa on orthorhombic $\alpha$-SnWO$_4$ have revealed that this compound undergoes two structural phase transitions within this pressure range. We have employed two distinct theoretical approaches, combined with experiments, to determine the crystal structure of the high-pressure phases, both of which are monoclinic. For the first structure, which we have named HP1 and is observed around 12.9 GPa, we have used a combination of DICVOL indexation and density-functional theory calculations together with a profile matching of X-ray diffraction patterns. In turn, the second high-pressure phase, named HP2 and detected around 17.5 GPa, has been identified using the CALYPSO code. The resulting monoclinic structure for HP2, which had not been previously reported in any orthotungstate compound, has been confirmed through a Le Bail fit to the experimental XRD patterns. The structural sequence observed in this work ($\alpha$->HP1->HP2) indicates two successive collapses of the unit-cell volume and an increase in the coordination number of Sn and W atoms. Our HP-XRD measurements and density-functional theory calculations have also allowed us to obtain the room-temperature equations of state and linear compressibility of the main axes for the three studied phases ($\alpha$-SnWO$_4$, HP1 and HP2). The data thus obtained indicate that HP2 exhibits a significant increase of the bulk modulus due to a remarkable compressibility reduction along one of its main compressibility axes. Finally, phonon dispersion and elastic constants have also been calculated to probe the dynamical and mechanical stability of the three SnWO$_4$ polymorphs. From the calculated elastic constants, the elastic moduli have been determined for the three phases. The values thus obtained indicate that, in the three polymorphs, shear deformation is easier than volumetric or uniaxial compression.

## CRediT authorship contribution statement

D.E. conceived the project. A.K. synthesized the sample. D.D.-A., R.O, J.I, P.B., F.A., and D.E. performed experiments, D.D.-A. and D.E. performed data analysis. L.W. and Y.L. performed CALYPSO calculations, A.M. carried out VASP calculations. All authors participated in discussions and writing and editing of the manuscript.

## Declaration of Competing Interest

The authors declare that they have no known competing financial interests or personal relationships that could have appeared to influence the work reported in this paper.

**Data Availability**

The data that support the findings of this study are available from the corresponding author upon reasonable request.

**Acknowledgments**

D.E. and A.M. gratefully acknowledge the financial support from the Spanish Research Agency (AEI) and Spanish Ministry of Science and Investigation (MCIN) under Projects PID2022-138076NB-C41/C44 and RED2022-134388-T (DOI: 10.13039/501100011033). D.E. would also like to thank the financial support of Generalitat Valenciana under grants PROMETEO CIPROM/2021/075-GREENMAT and MFA/2022/007. This study forms part of the Advanced Materials program and is supported by MCIN with funding from European Union Next Generation EU (PRTR-C17.I1) and by the Generalitat Valenciana. This study also was supported by the Foundation of Hebei Educational Committee in China (Grant No. BJK2023066). The authors thank Elettra synchrotron for providing beam time for the HP XRD experiments (Proposal 20230054).

**Appendix A. Supplementary information**

The supplementary information contains the crystal structure of phase HP3, HP4, HP5, and HP6 and the phonon dispersion of phases $\alpha$-SnWO$_4$, HP1, and HP2.

# References

[1] C. B. Liu, Z. Z. He, Y. J. Liu, R. Chen, M. M. Shi, H. P. Zhu, C. Dong, C. J. F. Wang, Magnetic anisotropy and spin-flop transition of NiWO₄ single crystals. J. Magn. Magn. Mat. 444 (2017) 190-192.

[2] O. Heyer, N. Hollmann, I. Klassen, S. Jodlauk, L. Bohatý, P. Becker, J. A. Mydosh, T. Lorenz, D. Khomskii, A new multiferroic material: MnWO₄. J. Phys. Cond. Matter 18 (2006) L471-L475.

[3] V. Dusastre, D. E. Williams, Selectivity and composition dependence of response of wolframite-based gas sensitive resistors (MWO₄)ₓ([Sn-Ti]O₂)₁₋ₓ (0< x< 1; M= Mn, Fe, Co, Ni, Cu, Zn). J. Mat. Chem. 9 (1999) 965-971.

[4] D. L. Stern, R. K. Grasselli, Propane oxydehydrogenation over metal tungstates. J. Catal. 167 (1997) 570-572.

[5] M. Dan, M. Cheng, H. Gao, H. Zheng, C. Feng, Synthesis and electrochemical properties of SnWO₄. J. Nanoscie. Nanotech. 14 (2014) 2395-2399.

[6] N. Sreeram, V. Aruna, R. Koutavarapu, D. Y. Lee, M. C. Rao, J. Shim, Fabrication of InVO₄/SnWO₄ heterostructured photocatalyst for efficient photocatalytic degradation of tetracycline under visible light. Envir. Res. 220 (2023) 115191.

[7] I. S. Cho, C. H. Kwak, D. W. Kim, S. Lee, K. S. Hong, Photophysical, photoelectrochemical, and photocatalytic properties of novel SnWO₄ oxide semiconductors with narrow band gaps. J. Phys. Chem. C 113 (2009) 10647-10653.

[8] D. Errandonea, J. Ruiz-Fuertes, A brief review of the effects of pressure on wolframite-type oxides. Crystals 8 (2018) 71.

[9] D. Errandonea, J. Pellicer-Porres, F. J. Manjón, A. Segura, C. Ferrer-Roca, R. S. Kumar, O. Tschauner, P. Rodríguez-Hernández, J. López-Solano, S. Radescu, A. Mujica, A. Muñoz, G. Aquilanti, High-pressure structural study of the scheelite tungstates CaWO₄ and SrWO₄. Phys. Rev. B 72 (2005) 174106.

[10] D. H. Fabini, G. Laurita, J. S. Bechtel, C. C. Stoumpos, H. A. Evans, A. G. Kontos, Y. S. Raptis, P. Falaras, A. Van der Ven, M. G. Kanatzidis, R. Seshadri, Dynamic stereochemical activity of the Sn²⁺ lone pair in perovskite CsSnBr₃. JACS 138 (2016) 11820-11832.

[11] W. Jeitschko, A. W. Sleight. Stannous tungstate: properties, crystal structure and relationship to ferroelectric SbTaO₄ type compounds. Acta Crystal. B 30 (1974) 2088-2094.

[12] W. Jeitschko, A. W. Sleight. Synthesis, properties and crystal structure of β-SnWO₄. Acta Crystal. B 28 (1972) 3174-3178.

[13] H. K. Mao, X. J. Chen, Y. Ding, B. Li, L. Wang, Solids, liquids, and gases under high pressure. Rev. Mod. Phys. 90 (2018) 015007.

[14] D. Errandonea, H. H. H. Osman, R. Turnbull, D. Diaz-Anichtchenko, A. Liang, J. Sanchez-Martin, C. Popescu, D. Jiang, H. Song, Y. Wang, F. J. Manjon, Pressure-induced hypercoordination of iodine and dimerization of I₂O₆H in strontium di-iodate hydrogen-iodate (Sr(IO₃)₂HIO₃). Mat. Today Adv. 22 (2024) 100495.

[15] J. Ruiz-Fuertes, S. López-Moreno, D. Errandonea, J. Pellicer-Porres, R. Lacomba-Perales, A. Segura, P. Rodríguez-Hernández, A. Muñoz, A.H. Romero, J. González, High-pressure phase transitions and compressibility of wolframite-type tungstates, J. Appl. Phys. 107 (2010) 083506.

[16] D. Errandonea, F. J. Manjon, Pressure effects on the structural and electronic properties of $ABX_4$ scintillating crystals. Prog. Mat. Scie. 53 (2008) 711-773.

[17] J. Ruiz-Fuertes, A. Friedrich, O. Gomis, D. Errandonea, W. Morgenroth, J. A. Sans, D. Santamaria-Perez, High-pressure structural phase transition in $MnWO_4$. Phys. Rev. B 91 (2015) 104109.

[18] D. Errandonea, F. Rodriguez, R. Vilaplana, D. Vie, S. Garg, B. Nayak, N. Garg, J. Singh, V. Kanchana, G. Vaitheeswaran, Band-Gap energy and electronic d–d transitions of $NiWO_4$ studied under high-pressure conditions. J. Phys. Chem. C 127 (2023) 15630-15640.

[19] E. Bandiello, P. Rodríguez-Hernández, A. Muñoz, M. B. Buenestado, C. Popescu, D. Errandonea, Electronic properties and high-pressure behavior of wolframite-type $CoWO_4$. Mat. Adv. 2 (2021) 5955-5966.

[20] A. Kuzmin, A. Anspoks, A. Kalinko, J. Timoshenko, R. Kalendarev, External pressure and composition effects on the atomic and electronic structure of $SnWO_4$. Sol. Energy Mat. and Sol. Cells 143 (2015) 627-634.

[21] A. Kuzmin, A. Anspoks, A. Kalinko, J. Timoshenko, R. Kalendarev, L. Nataf, F. Baudelet, T. Irifune, High-pressure x-ray absorption spectroscopy study of tin tungstates. Phys. Scripta 90 (2015) 094003.

[22] A. Kuzmin, A. Anspoks, A. Kalinko, J. Timoshenko, R. Kalendarev, L. Nataf, F. Baudelet, T. Irifune, P. Roy, Pressure-induced insulator-to-metal transition in $\alpha$-$SnWO_4$. J. Phys. Conf. Series 712 (2016) 012122.

[23] E. O. Gomes, A. F. Gouveia, L. Gracia, A. Lobato, J. M. Recio, J. Andrés, A chemical-pressure-induced phase transition controlled by lone electron pair activity. J. Phys. Chem. Letters 13 (2022) 9883-9888.

[24] R. M. Hazen, Temperature, pressure and composition: structurally analogous variables. Phys. Chem. of Min. 1 (1977) 83-94.

[25] D. Errandonea, J. Pellicer-Porres, F. J. Manjón, A. Segura, C. Ferrer-Roca, R. S. Kumar, O. Tschauner, J. López-Solano, P. Rodríguez-Hernández, S. Radescu, A. Mujica, A. Muñoz, G. Aquilanti, Determination of the high-pressure crystal structure of $BaWO_4$ and $PbWO_4$. Phys. Rev. B 73 (2002) 224103.

[26] J. L. Solis, J. Frantti, V. Lantto, L. Häggström, M. Wikner, Characterization of phase structures in semiconducting $SnWO_4$ powders by Mössbauer and Raman spectroscopies. Phys. Rev. B 57 (1998) 13491.

[27] C. Prescher, V. B. Prakapenka, DIOPTAS: a program for reduction of two-dimensional X-ray diffraction data and data exploration. High Pres. Res. 35 (2015) 223-230.

[28] S. Klotz, J. C. Chervin, P. Munsch, G. Le Marchand, Hydrostatic limits of 11 pressure transmitting media. J. Phys. D: Appl. Phys. 42 (2009) 075413.

[29] D. Errandonea, R. Turnbull, J. Sánchez-Martín, R. Oliva, A. Muñoz, S. Radescu, A. Mujica, L. Blackburn, N. C. Hyatt, C. Popescu, J. Ibáñez-Insa, A comparative study of the high-pressure structural stability of zirconolite materials for nuclear waste immobilisation, Results in Physics 61 (2024) 107704.

[30] A. Dewaele, P. Loubeyre, M. Mezouar, Equations of state of six metals above 94 GPa. Phys. Rev. B 70 (2004) 094112.

[31] R. O. Jones, Density functional theory: Its origins, rise to prominence, and future. Rev. Mod. Phys. 87 (2015) 897-923.

[32] P. E. Blöchl, Projector augmented-wave method. Phys. Rev. B 50 (1994) 17953-17979.

[33] G. Kresse, D. Joubert, From ultrasoft pseudopotentials to the projector augmented-wave method. Phys. Rev. B 59,(1999) 1758-1775.

[34] G. Kresse, J. Hafner, Ab initio molecular dynamics for liquid metals. Phys. Rev. B 47 (1993) 558-561.

[35] J. P. Perdew, A. Ruzsinsky, G. I. Csonka, O. A. Vydrov, G. E. Scuseria, L. A. Constantin, X. Zhou, K. Burke, Restoring the density-fradient expansion for exchange in solids and surfaces. Phys. Rev. Lett. 100 (2008) 136406.

[36] H. J. Monkhorst, J. D. Pack, Special points for Brillouin-zone integrations. Phys. Rev. B 13, (1976) 5188-5192.

[37] K. Parlinski, Computer Code PHONON, (2008). See: http://wolf.ifj.edu.pl/phonon.

[38] Y. Le Page, P. Saxe, Symmetry-general least-squares extraction of elastic data for strained materials from ab initio calculations of stress. Phys. Rev. B 65 (2002) 104104.

[39] Y. Wang, J. Lv, L. Zhu, Y. Ma, Crystal structure prediction via particle-swarm optimization. Phys. Rev. B 82 (2010) 094116.

[40] Y. Wang, J. Lv, L. Zhu, Y. Ma, CALYPSO: A method for crystal structure prediction. Comp. Phys. Commun. 183 (2012) 2063-2070.

[41] F. Birch, Finite elastic strain of cubic crystals. Phys. Rev. 71 (1947) 809-822.

[42] J. Gonzalez-Platas, M. Alvaro, F. Nestola, R. Angel, EosFit7-GUI: a new graphical user interface for equation of state calculations, analyses and teaching. J. Appl. Crystal. 49 (2016) 1377-1382.

[43] D. Diaz-Anichtchenko, J. E. Aviles-Coronado, S. López-Moreno, R. Turnbull, F. J. Manjón, C. Popescu, D. Errandonea, Electronic, vibrational, and structural properties of the natural mineral ferberite ($FeWO_4$): A high-pressure study. Inorg. Chem. 63 (2024) 6898-6908.

[44] I. Kawada, K. Kato, T. Fujita, $BaWO_4$-II (a high-pressure form). Acta Crystal. B 30 (1974) 2069-2071.

[45] O. Gomis, J. A. Sans, R. Lacomba-Perales, D. Errandonea, Y. Meng, J. C. Chervin, A. Polian, Complex high-pressure polymorphism of barium tungstate. Phys. Rev. B 86 (2012) 054121.

[46] M. J. Cliffe, A. L. Goodwin, PASCal: a principal axis strain calculator for thermal expansion and compressibility determination. J. Appl. Crystal. 45 (2012) 1321-1329.

[47] G. Grimvall, B. Magyari-Kope, V. Ozolin, K. A. Persson, Lattice instabilities in metallic elements. Rev. Mod. Phys. 84 (2012) 945-986.

[48] D. H. Chung, W. R. Buessem, The Voigt-Reuss-Hill Approximation and Elastic Moduli of Polycrystalline MgO, $CaF_2$, β-ZnS, ZnSe, and CdTe, J. Appl. Phys. 38 (1967) 2535–2540.

[49] T. Ouahrani, A. B. Garg, R. Rao, P. Rodríguez-Hernández, A. Muñoz, M. Badawi, D. Errandonea, High-pressure properties of wolframite-type $ScNbO_4$. J. Phys. Chem. C 126 (2022) 4664-4676.

[50] M. W. Stoltzfus, P. M. Woodward, R. Seshadri, J.-H. Klepeis, B. Bursten, Structure and bonding in $SnWO_4$, $PbWO_4$, and $BiVO_4$: Lone pairs vs inert pairs. Inorg. Chem. 46, (2007) 3839-3850.